\documentclass[letterpaper, 10 pt, conference]{ieeeconf}  % Comment this line out
\IEEEoverridecommandlockouts                              % This command is only
\usepackage{latexsym}
\usepackage{amsmath} % assumes amsmath package installed
\usepackage{amssymb}  % assumes amsmath package installed
\usepackage{amsxtra}
\usepackage{amsfonts}
\usepackage{graphics} % for pdf, bitmapped graphics files
\usepackage{epsfig} % for postscript graphics files
\usepackage{float}
\usepackage{graphicx}
\usepackage{setspace}
\usepackage{color}
\usepackage{subfigure}
\usepackage{psfrag}
\usepackage{dsfont}
\usepackage{multirow}
\usepackage{algorithm}
\usepackage{algorithmic}
\usepackage{authblk}
\newtheorem{theorem}{Theorem}[section]

\newcommand{\comment}[1]{}

\renewcommand{\baselinestretch}{1}

\title{\LARGE \bf Stability Analysis of a Model for the Market Dynamics of a Smart Grid}

\author[1]{F. Sorrentino\thanks{fsorrent@unm.edu}}
\author[2]{D. Toli\'{c}\thanks{domagoj.tolic@fer.hr}}
\author[3]{R. Fierro\thanks{rfierro@ece.unm.edu}}
\author[4]{S. Picozzi\thanks{sergio.picozzi@gmail.com}}
\author[1]{J. R. Gordon\thanks{jgordo01@unm.edu}}
\author[1]{A. Mammoli\thanks{mammoli@unm.edu}}
\affil[1]{Department of Mechanical Engineering, University of New Mexico, Albuquerque, NM, USA}
\affil[2]{Department of Control and Computer Engineering, University of Zagreb, Zagreb Croatia}
\affil[3]{Department of Electrical and Computer Engineering, University of New Mexico, Albuquerque, NM,  USA}
\affil[4]{Department of Physics, University of Maryland, College Park, MD, USA}

\begin{document}
\maketitle
\thispagestyle{empty}
\pagestyle{empty}

%%%%%%%%%%%%%%%%%%%%%%%%%%%%%%%%%%%%%%%%%%%%%%%%%%%%%%%%%%%%%%%%%%%%%%%%%%%%%%%%
\begin{abstract}
We consider the dynamics of a smart grid system characterized by widespread distributed generation and storage devices. We assume that agents are free to trade electric energy over the network  and we focus on the emerging market dynamics. We consider three different models %a simple model
for the market dynamics for which we present a stability analysis. We see that stability depends on the specific form of the market dynamics and it may depend on the structure of the underlying network topology.  We run numerical simulations that confirm our theoretical predictions.
As an example, we test our model for the market dynamics over a real network topology, namely, the Tramway 11 Feeder from New Mexico's power network.
\end{abstract}

\section{Introduction}

In this paper, we take a close look at the ongoing technological revolution that is transforming the power network into a \emph{smart grid} and %are particularly interested in modeling
we focus on its social and economical impact. %Our motivation is the observation of the current technological trends and our goal is to take a peek at how these are going to affect the society.
The fraction of electricity produced from renewables is expected to increase substantially over the years to come, in response to environmental concerns. This however poses a technological challenge as many renewable resources are intermittent by nature, with variations that may occur over a wide range of timescales, from seconds to hours,  which in addition may be difficult to predict. This variability may result in instabilities, poor power quality, and may limit the penetration of renewables.

A possible solution to these problems is the introduction of storage devices to complement intermittent power generators. Storage can serve
the dual purpose of compensating fast intermittency and delivering power when it is required rather than when it is available \cite{schoenung1996energy,hall2008energy}. Although
storage technologies are advancing rapidly, their cost can still be prohibitive, unless mechanisms are introduced that increase economic benefits
that can result from their introduction, for example real-time pricing or energy trading \cite{mammoli2012distributed}. Despite these benefits,
the potential drawbacks of allowing market transactions
to determine energy flows and associated costs are manifold. For example, fast decentralized dynamics enabled by automated computer-based
trading can lead to rapid fluctuations in prices and possible system collapse \cite{kirilenko2011flash,easley2011microstructure}. Moreover, although agents operate
on a local basis, it is possible that small but rapid local disruptions might eventually lead to large-scale cascading failures \cite{sterbenz2011evaluation}.

In this paper, we consider a situation where a large number of coupled agents control distributed energy resources (e.g. solar photovoltaic) sited in buildings, stores, factories, or as standalone facilities, operating under a variety of regulatory environments. We propose a simple dynamical model for the market dynamics of a smart grid. The main two features we incorporate in our model are the following: (i) devices for the storage of energy are available to a large number of agents;  (ii)  agents have broad access to a free market of energy which allows direct energy transactions between them. It has been suggested \cite{chassin2008decentralized} that (i) and (ii) may become common place practice in a near future smart grid, and lead to improvements over the status quo. We thus focus on variations on a simple model for the market dynamics for a smart grid that displays both features (i) and (ii).
%We consider a situation where local decentralized decisions are taken by the agents in order to: ensure satisfaction of their needs at any time and increase the efficiency of the process for the supply of energy.
Our goal is studying stability of the resulting complex dynamics.

There are several energy storage technologies which may be considered for the present study, including several types of battery,
as well as thermal storage. Battery storage includes a battery with a particular chemistry (e.g. lead-acid, lithium ion, flow battery) and an inverter, and may be characterized by energy capacity, power and round trip efficiency (i.e. the ratio of energy that can be extracted to the energy originally injected). Another form of energy storage, which serves the purpose of meeting the cooling load of a facility without a concurrent electric draw, is thermal storage. For the purposes of our model, thermal storage could just be considered as energy with no export value. The cost of battery technology is expected to decrease sharply in the future, while performance is expected to improve. This is the basis of our assumption of widely available storage capacity.

%On a practical note, there are several examples of facilities with distributed generation and/or storage. For example, the University %of
%New Mexico's Mechanical Engineering building is being used to demonstrate optimal scheduling of thermal storage and electric
%loads \cite{mammoli2010energetic}; a jail in Santa Rita, CA is combining PV generation with battery storage to reduce its operating %costs \cite{bossart2009renewable}; a distribution feeder
%in Albuquerque, NM, is hosting a number of distributed energy resources including a building-scale microgrid, a utility-scale PV %generation
%facility with battery storage for smoothing and load shifting \cite{ton2012us}.

We propose a simple dynamical description for the market dynamics, which we seek to analyze and we assume that the results of our analysis will capture some key-aspects of the dynamics when the number of agents becomes large enough. In our description, each individual agent is modeled as a dynamical unit, resulting in a high-dimensional system. %, where each agent is described as a dynamical unit.
In order to study stability, we will attempt a reduction of the stability problem in a low-dimensional form. Low-dimensional approaches have proved helpful in analyzing the dynamics of networks of coupled dynamical systems. Examples include (i) the stability of the synchronous evolution for both networks and hypernetworks of coupled oscillators \cite{FujiYama83,Replace,Pe:Ca,NSG,SAS,HYP,irving2012synchronization}, (ii) the stability of the consensus state in networks of coupled integrators \cite{CONS}, (iii) the stability of discrete state models of genetic control \cite{POM}, and (iv) the stability of strategies in networks of coupled agents playing a version of the prisoner's dilemma \cite{GT_SM}.

In particular, one of our motivations for studying stability of the market dynamics comes from the observation of the dynamics of the stock market, which is characterized by large temporal fluctuations, sometimes caused by automated trading algorithms \cite{kirilenko2011flash,easley2011microstructure}.
% known as \emph{speculative bubbles}.
We predict that if such instabilities were present in the energy market, the consequences might be even more serious than for the case of the stock market, namely disruptions to the real economy caused by blackouts and by large fluctuations in energy prices. While today's disruptions in electric power service are largely triggered by natural events or human errors, the type of instabilities we consider in this paper would result from the transition from a centralized control architecture to a decentralized one. The goal of this paper is to study the conditions under which a decentralized model of operations for the grid can be associated with stability of the market dynamics.

\section{Market models}

\begin{figure}[t]
\centering
\includegraphics[width=0.45\textwidth,trim=100pt 80pt 100pt 80pt]{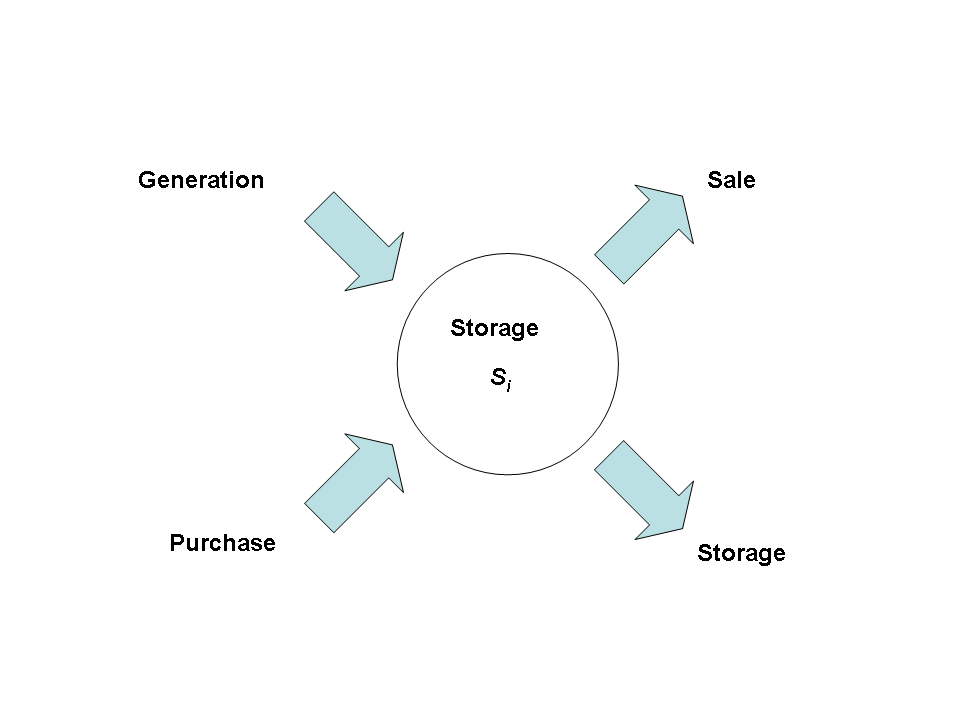}
\caption{The storage at each unit $i$ evolves based on both incoming flows (generation of electric energy and purchase of electric energy) and outgoing flows (consumption of electric energy and sale of electric energy). For the sake of this paper, we consider the net generation (generation minus consumption) as an exogenous uncontrollable variable (an input), while the dynamics of purchase and sale of electric energy between different network units is explicitly  modeled and studied.} \label{fig1}
\end{figure}

We consider a distribution-level smart grid of $N$ interacting agents/nodes. Each agent is allowed to freely trade energy according to  market rules. At each time $n$, the storage of electric energy at node $i$ is $S_i(n) \geq 0$, $i=1,...,N$. We assume that the agents are connected over a network.
The network connections are described by the entries of the $N$-square adjacency matrix $A=\{A_{ij}\}$, where $A_{ij}=A_{ji}=0$ if there is no direct connection from node $j$ to node $i$, while $A_{ij}=A_{ji}=1$ if there is a direct connection from node $j$ to node $i$, $A_{ii}=0$, $i=1,...,N$ (a direct connection is assumed to be one where power from meter $i$ flows to meter $j$ without passing through any other intervening meter). As we shall see, the constraint that the entries of the adjacency matrix $A$ are symmetric is a consequence of the conservation of energy for the transactions between different nodes (i.e., when a flow $A_{ij}(p_i-p_j)$ enters node $i$, the same flow leaves from node $j$. See also Fig.~\ref{fig:tanks}). Note
also that we are neglecting both electrical losses and storage losses for clarity and to simplify the analysis to its essential elements.

At each time $n$ and each node $i$, the storage of electric energy $S_i(n)$ can increase (decrease) if the generation exceeds the consumption of electric energy and  if the purchase exceeds the sale of electric energy. This is exemplified in Fig.\ 1. In this paper, we consider the \emph{net generation} (generation minus consumption) as an exogenous uncontrollable variable (an input), while the dynamics of purchase and sale of electric energy between different network units is explicitly modeled and its stability is investigated.

We consider that each agent sets a target storage $\bar{S}_i>0$, which, for simplicity, we assume to be constant with respect to time. This assumption can be retained as long as $\bar{S}_i$ evolves on a timescale that is much slower than that on which electric energy is produced or depleted. Moreover, as we shall see, none of our results explicitly depends on the particular choice of the targets $\bar{S}_i$, $i=1,...,N$.
At each time $n$, $S_i(n)$ can be either above or below $\bar{S}_i$; for example, some agents may have less storage of energy than they need (and therefore place a higher value on it), while others (presumably, fewer agents) may have more (valuing it less). Hence, it makes sense to consider a market in which agents are allowed to trade energy.

In what follows, we consider three different scenarios. In Scenario I, while agents are free to trade energy over the grid, we consider the existence of one special agent, namely, \emph{the utility}, which guarantees  availability of energy to the others. Each agent can either buy energy from the utility or from the other agents, in the case there are agents willing to sell it, at a presumably lower price. The lower price has to account for the fact that energy is sold based on availability and there is no obligation on these agents in guaranteeing a service to the others.
In Scenario II agents are allowed to trade energy over the grid, but they are all equal, i.e., there is no special entity that ensures that energy is always provided, if needed. More likely, \emph{the utility} or \emph{the utilities} are now agents like the others and free to either buy or sell energy based on convenience.
In Scenario III, we further analyze the dynamics of a smart grid whose agents are free to engage in market transactions, but we consider a fully developed market, i.e., for which at each time $n$ a unique \emph{market price} is set, based on the quantities demanded and supplied by all the individual agents.
%In scenarios IV and V, we will consider the formation of clusters of nodes, i.e., intermediate structures which allow to share resources among nodes and at the same time, to reduce the total complexity associated with the exchange of information over the network. While in  scenario IV we consider that clusters are isolated from each other, this constraint is removed in scenario V where
%cluster-to-cluster transactions are allowed over time-scales typically longer than for intra-cluster transactions.
%%situation in which energy is perceived as a rare commodity, whose price can quickly fluctuate in time. In this case, agents will have to adopt different strategies to cope with the fluctuations of the price of energy.

For all three cases above, we will be interested in evaluating stability of the dynamics about specific states, corresponding to specific well-behaved configurations of the system. All three scenarios are introduced and discussed.
 However, due to space limitations we mostly focus on analyzing Scenario I, while a thorough study of Scenarios II and III is left for future investigations. Note that in Scenarios II and III, the possibility that not all agents' energy needs can be satisfied, at any price, does exist.

\subsection{Scenario I}

We assume that agents are able to store energy and are free to trade energy over the grid. At each time step $n$,
$S_i(n)>\bar{S}_i$ for some agents, with the difference $(S_i(n)-\bar{S}_i)$ contributing to increase the total supply of energy, while $S_i(n)<\bar{S}_i$ for others, with the difference $(\bar{S}_i-S_i(n))$ contributing to increase the total demand for energy.
 However, there is no guarantee that the total offer of energy will correspond to the total
demand of energy (here we implicitly assume the total demand to be higher). The difference, i.e., the residual demand
 $D=\sum_{i=1}^N (\bar{S}_i-S_i)$ will have to be provided by the utility. We therefore assume that the price at which the utility sells energy, depends on the residual demand, according to the following relation,
\begin{equation}
p_u(n)=g_u(\sum_{i=1}^N (\bar{S}_i-S_i(n))),
\end{equation}
where the function $g_u(x)$ is an increasing function of its argument $x$.
Note that though we consider  market rules, %based on which agents are free to trade energy with other agents,
we still provide the utility a special role, i.e., the utility is a special agent in the system.

Transactions between agents are determined by a threshold price $p_i(n)$ set by each agent $i = 1,...,N$, above which agent $i$ is going to sell energy and below which it is going to buy energy. In other words, $p_i(n)$ represents a threshold that marks the transition between buying and selling mode for agent $i$.
We assume that if agent $i$ is willing to sell energy, it will fix a price $p_i(n)$ that is lower than the price set by the utility $p_u(n)$ and is determined by the relation,
\begin{equation}\label{due}
p_i(n)=p_u(n)-g_i(S_i(n)-\bar{S}_i),
\end{equation}
where the function $g_i(x)$ is a strictly increasing function of its argument $x$ ($g'_i(x)>0$), $g_i(0)=0$, $i=1,..,N$ and represents how the individual node $i$  reacts to variations in its storage in determining its price. Then the convenience for agents to buy energy from  other agents rather than from the utility derives from its lower price. At the same time, those agents who are local generators of energy and dispose of more energy than needed will be willing to sell it for a lower price. %Though it cannot be excluded that an agent will purchase energy from the utility at a lower price and may sell it at a higher price at a later time, we will focus on the simple case that agents become sellers when their storage of energy exceeds their needs.

\begin{figure}[t]
\centering
\includegraphics[width=3.5in]{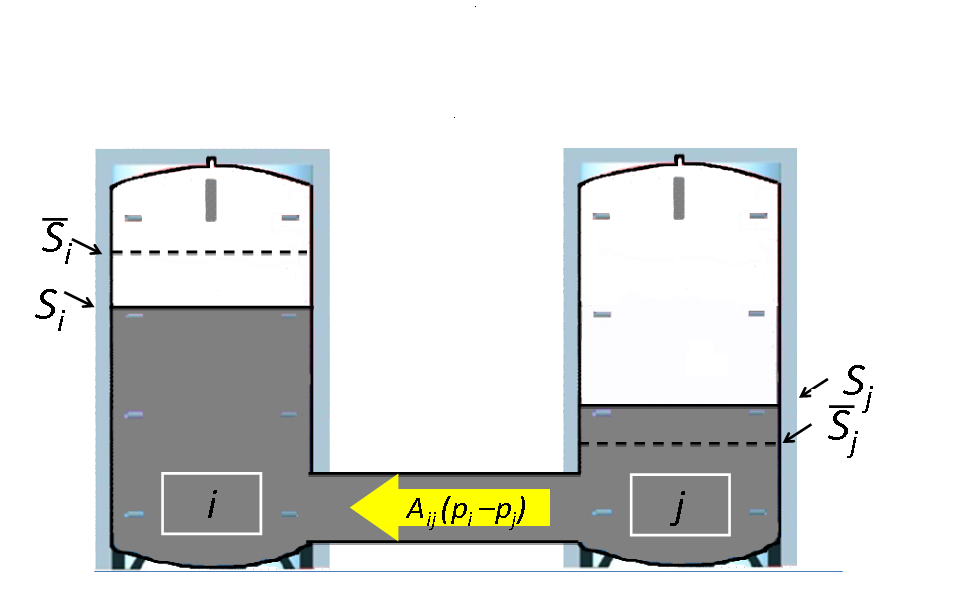}
\caption{The figure exemplifies our scheme for the evolution of the storages. For each agent $i$, $S_i$ is the current storage, $\bar{S}_i$ is the target storage, and $p_i$ in Eq.\ (\ref{due}) represents the price above which $i$ behaves as a seller and below which it behaves as a buyer. Flows are determined by the prices set by the nodes. Given two nodes $i$ and $j$, with  $p_i>p_j$, energy will flow from $j$ to $i$, even when $S_i>S_j$ (case shown in the figure). The flow of energy from $j$ to $i$ will be equal to $A_{ij}(p_i-p_j)$.}
\label{fig:tanks}
\end{figure}

Through an analogy with an electric network, we can imagine that $p_i(n)$ represents something equivalent to the electric potential of the node $i$: as electrons would flow from the nodes at low electric potential to those at high electric potential, in our case,  energy will flow from the nodes for which $p_i$ is low to those for which $p_i$ is high. This is illustrated in Fig.~\ref{fig:tanks}. Flows are only allowed over those links that actually exist in the network, i.e., for which $A_{ij}=A_{ji}>0$. Hence,
\begin{equation}\label{A}
S_i(n+1)=S_i(n)+\sum_j A_{ij} [p_i(n)-p_j(n)]+u_i(n),
\end{equation}
where $u_i(n)$ represents the net generation/consumption of energy at node $i$ at time $n$. A net generation (consumption) corresponds to a positive (negative) $u$.
By using Eqs. (\ref{due}) and (\ref{A}), we obtain
%\begin{subequations}
\begin{align} \label{main}
S_i(n+1)= & S_i(n)   +  \sum_j A_{ij} [g_j(S_j(n)-\bar{S}_j) \\ & -g_i(S_i(n)-\bar{S}_i)]  +u_i(n) \nonumber \\ = & S_i(n)+\sum_j L_{ij} g_j(S_j(n)-\bar{S}_j)+u_i(n),
\end{align}
%\end{subequations}
 where the Laplacian matrix $L=\{L_{ij}\}$ is such that $L_{ij}= (A_{ij} -\delta_{ij} \sum_k A_{ik}$), $i,j=1,...,N$, and $\delta_{ij}$ is the Kronecker delta. By the assumption that the matrix $A$ is symmetric, $L$ is also symmetric and its spectrum is real.
 The matrix $L$ is negative semidefinite, hence all of its eigenvalues $\{\lambda_i\}_{i=1}^N$ are nonpositive. By construction, one eigenvalue, say $\lambda_1$ is equal zero, with associated right eigenvector $[1,1,...,1]^T$.  Moreover, if the matrix $A$ is irreducible, i.e., if the underlying network is connected, then $\lambda_2,\lambda_3,...,\lambda_N$ are strictly negative; in what follows we proceed under the assumption that the matrix $A$ is irreducible.

The function of the storages $S_i$ is compensating for fast (possibly unexpected) temporal variations in the net generations/consumptions $u_i$'s. These in fact reflect changes in both the consumption and the generation of electric energy for each individual unit. Hence, over the timescale on which the $S_i$'s evolve, the $u_i$ terms can be considered as a bounded disturbance.

%{\bf IN THIS PARAGRAPH I WAS ASSUMING THAT THE $u_i=0$. I AM GOING TO REWRITE IT ASSUMING THE $u_i$ TO BE SMAlL.}
%
%As a first case, we consider that at any time the $u_i(n)$ are small quantities. We can compute an average storage for all the nodes in the network
%\begin{equation}
%S_{av}(n)=N^{-1} \sum_{j=1}^N S_j(n).
%\end{equation}
%We can then define deviations from this average storage
%\begin{equation}
%\delta S_i(n)=(S_i(n)-S_{av}(n))
%\end{equation}
%We are interested in studying the dynamics of the deviations $\delta S_i(n)$, and in particular whether they remain small or become large. We proceed under the assumption that $g_i(x)=g_*(x)$ for $i=1,...,N$. Then we linearize about the average storage $S_{av}(n)$ and obtain,
% \begin{equation} \label{lin}
% \delta S_i(n+1)= \delta S_i(n)+ \sum_j L_{ij} g'_j(0)\delta S_j(n),
% \end{equation}
% $i=1,...,N$, where $g'_j(x)\geq 0$ represents the derivative of the function $g_j(x)$ with respect to its argument $x$.
%{\bf I NEED TO WORK ON THIS}

We first perform a local stability analysis for the unperturbed dynamics, i.e., for the case that $u_i(t)=0$ in Eq. (\ref{main}), $i=1,...,N$. Successively, we extend our stability analysis to the case when $u_i (t)$ is unknown but bounded.
 In first approximation, we neglect the $u_i$ and compute the fixed points for the dynamics, Eq. (\ref{main}). If these fixed points turn out to be stable and the $u_i$'s are small, we expect fluctuations of the systems' trajectories about them, with the size of these fluctuations being determined by the $u_i$'s. On the other hand, if the fixed points are unstable, we expect the systems' trajectories to rapidly move away from them.  %Moreover, in this case,  the presence of the $u_i$'s ensures that the system does not remain in %the unstable fixed point configuration.
Then, we can obtain fixed points of the dynamics, Eq.\ (\ref{main}),  if the vector
\begin{eqnarray}
{\small\small\small{ \begin{pmatrix}
g_1(S_1(n)-\bar{S}_1) \cr
g_2(S_2(n)-\bar{S}_2)  \cr
... \cr
g_N(S_N(n)-\bar{S}_N)
\end{pmatrix}}}. \nonumber
\end{eqnarray}
belongs to the null subspace of the matrix $L$.  If we assume $g_i(x)=g_*(x)$ for $i=1,...,N$, then a fixed point of the dynamics will be of the type
\begin{equation}\label{ten_}
S_i=\bar{S}_i+z,
\end{equation}
$i=1,...,N$, with $z \in R$. Equation (\ref{ten_}) corresponds to all the storages $S_i$ deviating from their target $\bar{S}_i$, by the same quantity $z$ for $i=1,...,N$ and defines a line in phase space that goes through the point $(\bar{S}_1,\bar{S}_2,...,\bar{S}_N)$ and is parallel to the vector $[1,1,...,1]$. %Convergence on the points of this line depends on the choice of the initial conditions, as it can be proven that the quantity
%\begin{equation} \label{Q_}
% Q=\sum_j S_j(n)
% \end{equation}
%is an invariant of the dynamics. Hence, assuming stability, the final point will be obtained by imposing simultaneous satisfaction of conditions (\ref{ten_}) and (\ref{Q_}).

One particular solution is the point
 \begin{equation} \label{target}
 S_j(n)=\bar{S}_j,
 \end{equation}
 $j=1,...,N$, corresponding to setting $z=0$ in (\ref{ten_}), for which each unit storage is equal to the target storage for that unit.

\begin{theorem}\label{T}
We assume that all the nodes $j=1,...,N$ are characterized by the same function $g_j$, i.e., $g_1(x)=g_2(x)=...=g_N(x)=g_*(x)$. Then a necessary and sufficient condition for linear stability of (\ref{ten_}) is that the following two conditions are met: (I) $g'_*(0)>0$ and (II)  $M=\max_{j=2}^N |\lambda_j|< 2 g'_*(0)^{-1}$, where $|x|$ is the absolute value of $x$.
\end{theorem}

 By linearizing Eq. (\ref{main}) about the line (\ref{ten_}), we obtain,
 \begin{equation} \label{lin}
 \delta S_i(n+1)= \delta S_i(n)+ \sum_j L_{ij} g'_j(0)\delta S_j(n),
 \end{equation}
 $i=1,...,N$, where $g'_j(x)\geq 0$ represents the derivative of the function $g_j(x)$ with respect to its argument $x$. We are interested in determining whether the above small perturbations grow or decay back to the line (\ref{ten_}), in which case we will define the system to be stable.

 If we assume that all the nodes $j=1,...,N$ are characterized by the same function $g_j$, i.e., $g_1(x)=g_2(x)=...=g_N(x)=g_*(x)$, then Eq. (\ref{lin}) can be rewritten,
 \begin{equation}
 \delta S(n+1)=  (I+ g'_*(0) L )\delta S(n),
 \end{equation}
 where the column vector $\delta S(n)=[\delta S_1(n),\delta S_2(n),...,\delta S_N(n)]^T$. Hence, stability is achieved if the following condition holds,
 \begin{equation}
 |1+ g'_*(0) \lambda_i|<1,
 \end{equation}
$i=2,...,N$, where we have indicated with $|x|$ the absolute value of the number $x$. We observe that the eigenvalue $\lambda_1=0$ does not contribute to stability as it is associated with perturbations along the  line  that goes through the point $(\bar{S}_1,\bar{S}_2,...,\bar{S}_N)$ and is parallel to the vector $[1,1,...,1]$. Note that the system is neutrally stable with respect to such perturbations. % fall inside the circle of unit radius centered at $(0,0)$ in the complex plane. It suffices that only one eigenvalue lies outside of the unit circle to make the fixed point unstable.
This proves Theorem \ref{T}. Condition (II) highlights the effect of the network topology on stability, which depends on $M$ the maximum absolute value of the eigenvalues of the Laplacian matrix.

\begin{figure}[t]
\centering
\includegraphics[width=3.5in]{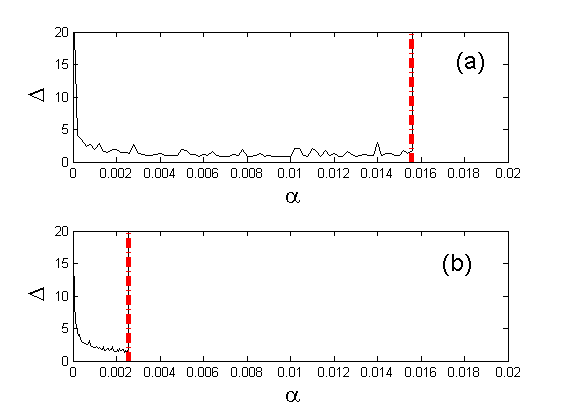}
\caption{We run numerical simulations for the set of equations (\ref{main}), $g_i(x)= \alpha x, i = 1,...,N$. Plot (a) shows the final difference $\Delta$ versus $\alpha$ for the case that the underlying network topology is an $N=1000$-random graph, $A_{ij} = 1$ with probability $0.1$ for $j \neq i$, $A_{ij} = 0$ otherwise. Plot (b) shows $\Delta$ versus $\alpha$ for a case in which we repeat the same numerical simulation but we consider a scale free network  \cite{korea}, with the same number of nodes $N=1000$ and edges as in (a). %As can be seen the underlying network topology has a strong effect on stability.
In both cases (a) and (b) $\alpha=0$ is unstable as condition (I) requires $\alpha>0$. The vertical dashed-dotted lines represent the threshold for instability predicted by our low-dimensional analysis (condition (II)).}
\label{Diff}
\end{figure}

Let us now consider the general case of perturbed dynamics by relaxing
the assumption that $u_i(n) = 0$ in (\ref{main}). In the following, we show
that under some reasonable assumptions the system is stable. To be more
specific, we assume that the vector $u(n)=[u_1(n),u_2(n),...,u_N(n)]^T$ is bounded by a small number
$\varepsilon_{1}$, and the initial condition $\|\delta S(0)\|<c_{1}$. Thus
Eq. (\ref{main}), after linearization, can be written as follows
\begin{equation}
 \delta S(n+1)=   (I+ g'_*(0) L ) \delta S(n) + u(n).
 \end{equation}
For the nominal system, ($u_i(n) = 0$), the line (6) is locally asymptotically stable
provided that conditions (I) and (II) are satisfied. Since $\|u\|\leq \varepsilon_1$ by using
stability
theory of perturbed systems \cite{Khalil+book:02}, and the condition that
$\left\|\delta S(0)\right\|<c_{1}$ for some
positive constant $c_{1}$, then
\begin{displaymath}
\left\|\delta S(n)\right\|\leq \sigma_{1},\qquad \forall\: n \geq n_{*}
\end{displaymath}
for some finite time $n_{*}$.

 We run two numerical experiments for the set of equations (\ref{main}), $g_i(x)= \alpha x, i = 1,...,N$; we first consider a network of $N=1000$ nodes whose adjacency matrix $A=\{A_{ij}\}$, is such that $A_{ij}$ is set to be  equal $1$ with probability $0.1$ for $j \neq i$, $A_{ij} = 0$ otherwise. The largest eigenvalue in modulus for this network is equal to $M=128.41$. Based on our analysis, we expect that a necessary and sufficient condition for stability is that (I) $\alpha>0$ and (II) $\alpha< 2/128.41 \simeq 0.0156$. The latter condition has a simple physical interpretation in terms of the rate  $\alpha$ that should not exceed a critical value in order to preserve stability, where this critical value is determined by the network topology. We iterate the set of equations (\ref{main}) from an initial condition $S_i(0)=\bar{S}_i+\varepsilon_i$, where we set the target storages $\bar{S}_i$ to be uniformly distributed random numbers between $0$ and $100$ and the $\varepsilon_i$ to be random numbers drawn from a normal distribution.
 For each node $i$ and each time $n$, the inputs $u_i(n)$ are random numbers drawn from a normal distribution. In Fig.\ 3(a) we plot the final difference
\begin{equation}\label{delta}
\Delta={(N(n_2-n_1))}^{-1} \sum_{n=n_1}^{n_2} \sum_{i=1}^N |S_i(n)-\bar{S}_i|,
\end{equation}
versus $\alpha$, where $n_1=900$ and $n_2=1000$. We have repeated the same numerical experiment for a case in which the random network is replaced by a scale free network generated by using the algorithm in \cite{korea}, with the same number of nodes $N=1000$ and edges
as before. The results of this simulation are shown in Fig.\ 3(b). For this case the largest eigenvalue in modulus is equal to $M=780.03$ and condition (II) becomes $\alpha< 2/780.03 \simeq 0.0026$. As can be seen by comparing plots (a) and (b), the underlying network topology has a considerable effect on stability. This also suggests the importance of designing the underlying network in order to provide improved stability.

  %and we observe that for small values of $\alpha$ there is a dramatic level of instability, then the system is well behaved for some values of $\alpha$ until it reaches a critical $\alpha$ and the system becomes unstable for any larger value.
  We note that when defining the storages variables $S_i$, we have only specified that these are positive quantities over an arbitrary reference system, $S_i \geq 0$. In any practical situation, we would also have to consider a maximum capacity for $S_i$, i.e., $0 \leq S_i \leq S_i^{max}$, where $S_i^{max}$ are given constants, $i=1,..,N$.  While this modification would have an effect on the dynamics described by Eqs. (\ref{main}), it would not modify our stability analysis, provided that the target storages $\bar{S}_i$ are set to be $0<\bar{S}_i< S_i^{max}$. In fact, under this condition, we would still be able to study the dynamics of infinitesimally small perturbations about $\bar{S}_i$, and the stability analysis would therefore remain unchanged.%We also note that as the storages $S_i$'s are defined over an arbitrary reference system, if at any time while iterating Eqs. (\ref{main}) $S_i$ became negative, we would be free to change the reference system so as to preserve the property that $S_i \geq 0$.

\begin{figure}[t]
\centering
\includegraphics[width=0.4\textwidth,trim=10pt 10pt 10pt 10pt]{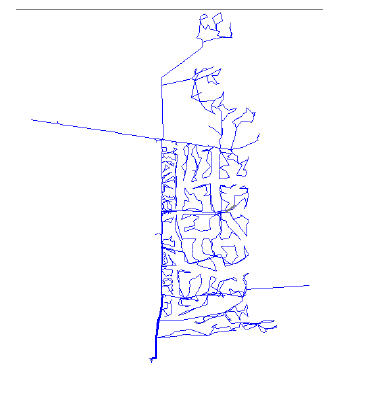}
\caption{Tramway 11 feeder from New Mexico's power network. The network comprises a total of 3806 nodes of which 1612 nodes represent physical loads in the network.}
\end{figure}

We also consider real data from the Tramway 11 feeder from New Mexico's power network (shown in Fig.\ 4). The datasets that we use have planar spatial coordinates for all the nodes in the network (including transformers and buses) and the physical connections between them.
Of the 3806 nodes, only 1612 nodes represent physical loads in the network. We consider the physical distance along the shortest paths between any pair of these load bearing nodes and consider two such nodes to be connected if the distance along the shortest path is less than a threshold $\theta$. Figure 5 shows the maximum  modulus for the eigenvalues of the corresponding Laplacian matrix versus $\theta$. As can be seen, $M$ increases with $\theta$ and saturates for large enough values of $\theta$ ($\theta$ approximately larger than $1.8 \times 10^4$), for which the network becomes fully connected.

We also test our model for the market dynamics on the network generated from the Tramway 11 feeder data with $\theta=10^3$. The corresponding maximum eigenvalue $M=49.22$. Based on our analysis, we expect that a necessary and sufficient condition for stability is that (I) $\alpha>0$ and (II) $\alpha< 2/49.22 \simeq 0.0406$.  For each node $i$ and each time $n$, the inputs $u_i(n)$ are random numbers drawn from a normal distribution. In Fig.\ 6 we plot the final difference $\Delta$
versus $\alpha$. As can be seen, the numerical simulations confirm our theoretical predictions about stability.

\begin{figure}[t]
\centering
\includegraphics[width=3.5in]{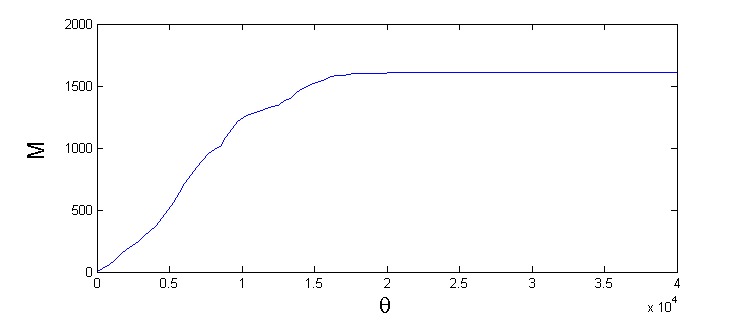}
\caption{Maximum modulus for the eigenvalues of the Laplacian matrix for the tramway 11 feeder from New Mexico's power network versus the threshold $\theta$ (see the main text for more details).}
\end{figure}

\begin{figure}[t]
\centering
\includegraphics[width=3.5in]{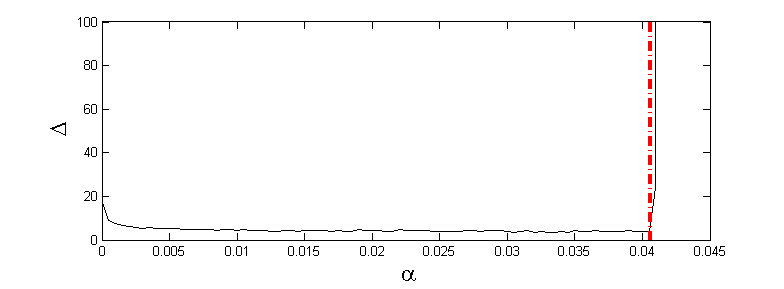}
\caption{We run numerical simulations for the set of equations (\ref{main}), $g_i(x)= \alpha x, i = 1,...,N$. The plot shows the final difference $\Delta$ versus $\alpha$ for the case that the underlying network topology is the tramway 11 feeder from New Mexico's power network, $\theta=10^3$. }
\end{figure}

\subsection{Scenario II: an alternative model that incorporates the dynamics of the prices}

In this section, we consider a scenario where all the agents in the network are free to trade energy but there is no utility to guarantee that energy is made available, if needed. We assume that the utility is now an agent like the others, i.e., it is free to buy or sell energy, based on convenience. Moreover, we assume that each agent has the freedom to update $p_i(n)$, defined as the price above which it is willing to sell energy and below which he/she is willing to buy energy. We are interested in describing the temporal dynamics of $p_i(n)$ for $i=1,...,N$. We expect that at each time step, the price will vary based on the difference  $(S_i(n)-\bar{S}_i)$, i.e., on how much the storage deviates from its target. If for example, at a given time step, the storage $S_i$ of node $i$ is below its target $\bar{S}_i$, $i$ will increase the price at which it is willing to buy energy from the grid. If at the following time step, $S_i$ is still less than $\bar{S}_i$, the price will rise again and so on.
%$p_i(n+1)$ will be a function of (i) $p_i(n)$, (ii) $p_u(n)$, (iii) $S_i(n)$.
A simple description of the dynamics is then the following,
\begin{equation} \label{pi}
p_i(n+1)=p_i(n)- \beta_i (S_i(n)-\bar{S}_i),
\end{equation}
where $\beta_i \geqslant 0$, $i=1,...,N$, together with Eq. (3) describing the evolution of the storages $S_i(n)$. Fixed points for the set of equations (3) and (\ref{pi}) are obtained if condition (\ref{target}) is verified (recall that by definition $g_i(0)=0$).
 %$j=1,...,N$, corresponding to setting $z=0$ in (\ref{ten}),
 This corresponds to setting the storage of each unit equal to the target storage for that unit. By linearizing Eqs. (3) and (\ref{pi}) about (\ref{target}), we obtain,
 \begin{subequations}\label{S2l}
 \begin{align}
 \delta S_i(n+1)= &\delta S_i(n)+ \sum_j L_{ij} \delta p_j(n), \\
 \delta p_i(n+1)= & \delta p_i(n)-\beta_i \delta S_i(n).
 \end{align}
 \end{subequations}
We look now for a solution of the linearized system (\ref{S2l}) that does not depend explicitly on the index $i$. To this end, we replace $\sum_j L_{ij} \delta p_j(n)$ by $\lambda_k \delta p_i(n)$, where $\lambda_k$ is an eigenvalue of the Laplacian matrix $k=1,...,N$ \cite{GT_SM}. With this substitution, we obtain
\begin{align}
 \delta S_i(n+1)= & \delta S_i(n)+ \lambda_k \delta p_i(n), \\
 \delta p_i(n+1)= & \delta p_i(n)-\beta_i \delta S_i(n),
 \end{align}
$i,k=1,...,N$. Stability depends on the eigenvalues of the matrix, $M_{ik}=$
\begin{eqnarray}
{\small\small\small{ \begin{pmatrix}
1  & \lambda_k \cr
-\beta_i & 1  \cr
\end{pmatrix}}}. \nonumber
\end{eqnarray}
In particular, for stability, the eigenvalues of this matrix have to be in modulus less than one, for $i,k=1,...,N$. However an important observation is that the eigenvalues of the matrix $M_{ik}$ are equal to $1\pm\sqrt{\lambda_k \beta_i}$, hence there will always be eigenvalues of modulus equal or larger than 1 for any choice of the pairs ($\lambda_k,\beta_i$). This points out the inherent instability of the model described  by Eqs. (3) and (\ref{pi}).

%To observe this behavior a random network of 500 nodes is generated and tested against the rules of this scenario. Taking the maximum eigenvalue when varying $\beta_i$ and $\lambda_k$ results in the plot below, which demonstrates the unstable eigenvalues for this scenario,
%%Considering $\lambda_k\leq0$,  a sufficient condition for stability is that
%%\begin{equation}
%%[(1+\alpha_{ik})^2+\beta_{ik}^2]<1/4,
%%\end{equation}
%%$i,k=1,...,N$, where $\alpha_{ik}+j \beta_{ik}=\sqrt{-g_i \lambda_k}$. In the case that the eigenvalues are real this condition reduces to $\frac{-2}{\alpha} \leq \lambda_{min} \leq 0$, %where $\lambda_{max}$ is the smallest possible eigenvalue of $L$ without seeing instability from the solutions.
%
%\begin{figure}[h]
%\centering
%\includegraphics[width=4in]{eigens.eps}
%\caption{We set a random network of $500$ nodes and plot the maximum eigenvalue of the matrix $M_{ik}$ versus $g$. }
%\label{Stab}
%\end{figure}

\subsection{Scenario III: fully developed market}

We have seen that Scenario II is always unstable. This motivates us to look into alternative forms of market, for which, given an appropriate choice of the parameters, it is possible to conciliate the dynamics of the prices with the stability of the network. %Moreover, we will try to describe , i.e., at each time each agent $i$ sets a price at which it seeks to either buy or sell a certain quantity of the demanded or supplied good, but

In a fully developed market, agents are given a chance to bid on a commodity, i.e., they can set a quantity that they are seeking to either buy or sell and a price at which they want the transaction to occur. However, there is no guarantee that the whole quantity demanded or supplied will be traded for that price, not even that the transaction will occur at all. Therefore, in what follows we will try to consider the uncertainty that typically characterizes market transactions.

 We consider that at each time $n$, the quantity to be either demanded or supplied by agent $i$ is given by,
\begin{equation}
q_i(n)=(S_i(n)-\bar{S}_i),
\end{equation}
where a positive $q_i$ is supplied and a negative $q_i$ is demanded.
At each time step $n$, \emph{the current market price} $p(n)$ for that commodity  is determined  from the %intersection of the
demand and supply curves. At that price usually corresponds a quantity of the market transaction equal to $q(n)$. %For example, one way to obtain $p(n)$ is to compute that price that %makes the total amount of sold energy equal to the total amount of bought energy,
%maximizes the amount of  transactions, i.e.,

Thus in general we can write the current market price $p(n)$ and the market quantity $q(n)$ as
\begin{subequations}
\begin{align}
p(n)= & F_p( p_1(n), q_1(n),p_2(n), q_2(n),...,p_N(n), q_N(n)),\label{price} \\
q(n)= & F_q( p_1(n), q_1(n),p_2(n), q_2(n),...,p_N(n), q_N(n)),
\end{align}
\end{subequations}
where $F_p$ and $F_q$ are functions to be determined. For example, specific forms for the functions $F_p$ and $F_q$  are associated with the algorithms commonly used in the stock market %in the financial markets
to determine the market price and the market quantity.
%[CAN WE GET AN IDEA OF HOW THESE ALGORITHMS ACTUALLY WORK?].
It is important to point out that, different from Scenario II, Eq.\ (\ref{price}) corresponds to assuming that the time-scale over which the price $p(n)$ is computed is much shorter than that on which the storage of energy   $S_i(n)$ evolves, thus $p(n)$ converges instantaneously at a unique equilibrium value for all the agents. In practice, this  requires that the market price is computed in a centralized way and communicated on all the agents.

It is important to consider the uncertainty that typically characterizes this type of transactions. At each time, each agent sets a price $p_i(n)$ and a quantity demanded/supplied $q_i(n)$. However, it is difficult for each individual agent $i$ to predict the market price at which the transaction will actually occur and the corresponding market quantity, as they also depend on the other agents' actions. This will affect the extent to which agent $i$ will be able to conclude the transaction.
% For example,  for the case of the financial markets, even if $p_i>p$ ($p_i<p$), it is still possible that agent $i$ will not be able to purchase (sell) for that price if there is a shortage of supply (demand). This usually results in the agents being willing to pay more (receive less) for the commodity they seek to buy (sell) to leverage against the probability of not concluding the transaction.

Consider for example a buyer $i$, demanding at time $n$ a quantity equal to $(\bar{S}_i-S_i(n))$. We introduce a coefficient $0 \leq \mathcal{C}_i(p_i,q_i) \leq 1$, indicating the fraction of the demanded quantity by $i$ that agent $i$ will be actually able to get. For example, it is to be expected that in the limit in which $i$ is willing to pay a very high (low) price, $\mathcal{C}_i \rightarrow 1$ ($\mathcal{C}_i \rightarrow 0$). For intermediate situations, we expect the fraction of the demanded quantity that $i$ is able to get to increase with $(p_i(n)-p(n))$.
%Different from $p(n+1)$, the current price $p(n)$ is an available information, and can be used to estimate $p(n+1)$ under the assumption that the market price will not vary much in one time step.
%To incorporate this effect, we introduce the probability of success $0 \leq \mathcal{C}_i \leq 1$ for agent $i$, i.e., theprobability that agent $i$ will be able to conclude the transaction, which we set to be an increasing function of $|p_i(n)-p(n)|$.
 In what follows, we choose $\mathcal{C}_i$ to be modeled by the following function,
\begin{equation}\label{Pi}
\mathcal{C}_i(n)=\tanh^2(c(p_i(n)-p(n))),
\end{equation}
$i=1,...,N$, where $c$ is a multiplicative constant.
Eq.\ (\ref{Pi}) indicates that setting a price well above or well below the market price will increase $\mathcal{C}_i$.

We note  that the relation (\ref{Pi}) does not take into account the demanded/supplied quantity $q_i(n)$ in determining $\mathcal{C}_i$. Using again the example of the buyer $i$, it is to be expected that the fraction of the demanded quantity $i$ will receive, will also depend on the amount of the demanded quantity, i.e., $\mathcal{C}_i$ should decrease with $|q_i|$ (and in fact, when the demand of a commodity is high, the price typically goes up). In order to take into account the effect of $q_i$ on $\mathcal{C}_i$, we propose a modified version of Eq. (\ref{Pi}),
%is clearly an approximation, as in reality the probability of concluding a transaction also depends on the quantity demanded or supplied. However, we deem that Eq.\ (\ref{Pi}) provides a reasonable approximation if the quantities demanded and supplied do not vary considerably from agent to agent. If one wanted to include also the effect of the quantities demanded/supplied on $\mathcal{P}_i$, (\ref{Pi}) could be rewritten,
\begin{equation}\label{Pi2}
\mathcal{C}_i(n)=\tanh^2\Bigg( \frac{c (p_i(n)-p(n))}{1+q_i(n)^2}\Bigg),
\end{equation}
$i=1,...,N$. Eq.\ (\ref{Pi2}) indicates that large (small) demanded/supplied quantities will have the effect of decreasing (increasing) $\mathcal{C}_i$. %For the sake of simplicity, in what follows we will proceed under the assumption that $\mathcal{P}_i$ is  described by (\ref{Pi}).

Having introduced the quantity $\mathcal{C}_i$, Eq. (\ref{A}) can be modified as follows,
\begin{equation}\label{S3a}
S_i(n+1)=S_i(n)- \mathcal{C}_i(n) (S_i(n)-\bar{S_i})+u_i(n),
\end{equation}
$i=1,...,N$, which for $\mathcal{C}_i=1$ yields $S_i(n)=\bar{S}_i+ u_i(n)$ and for $\mathcal{C}_i=0$ yields $S_i(n)={S}_i(0)+ \sum_{n'=0}^n u_i(n')$.
%where  $S_i(n)$ now represents the expected storage of energy at node $i$ at time $n$. For probability of success $\mathcal{P}_i=1$, we have that $S_i=\bar{S}_i+\sum_{n'=0}^n u_i(n')$.
Moreover, as in Scenario II, we assume that
\begin{equation}\label{S3b}
p_i(n+1)=p_i(n)- \beta_i(S_i(n)-\bar{S}_i),
\end{equation}
$i=1,...,N$, where $\beta_i>0$.
We see that a fixed point of the dynamics (described by Eqs.\ (\ref{S3a}) and (\ref{S3b})) is given by $S_i=\bar{S}_i$, $p_i=p_i^*$, $i=1,...,N$, and $p=p^*$, where $p^*$ is determined by setting $p_i=p_i^*$ and $q_i=0$ for $i=1,..,N$ in Eq.\ (\ref{price}).   By linearizing Eqs. (\ref{S3a}) and (\ref{S3b}) about this fixed point, we obtain,
\begin{subequations} \label{Sl3}
\begin{align}
\delta S_i(n+1)=& [1 -   \tanh^2(c(p_i^*-p^*))] \delta S_i(n),\\
\delta p_i(n+1)=& \delta p_i(n)- \beta_i \delta S_i(n).
\end{align}
\end{subequations}
Note that Eq.\ (\ref{Sl3}) is independent of whether we chose $\mathcal{C}_i$ to be described by  (19) or (20).
 As the function $0 \leq \tanh^2(x)<1$, it follows that the system (\ref{Sl3}) is always neutrally stable in the sense that the perturbations $\delta S_i(n) \rightarrow 0$, while the perturbations $\delta p_i$ neither grow nor shrink. Interestingly, determination of whether the fixed point is stable or not is not affected by the particular choice of the function $p(n)$ in (\ref{price}). We conclude that this scenario is stable, in the sense that $S_i \rightarrow \bar{S}_i$ for $i=1,...,N$.

We wish to emphasize that we have been able to resolve the instability of the market dynamics observed for Scenario II by considering an alternative model for which (i) at each time $n$ a unique market price $p(n)$ and market quantity $q(n)$ are given (ii)  the extent to which agent $i$ is able to either buy or sell a demanded quantity  is determined by an appropriately defined coefficient $\mathcal{C}_i$.

%Then it can be seen that stability is achieved if $-2 < -\tanh^2(p_i^*-p^*) < 0$, $i=1,...,N$, corresponding to a bound on the maximum allowed variation of the price $p_i$ set by each individual agent with respect to the market price $p$.

\section{CONCLUSION}

We have considered the dynamics of a smart grid system characterized by widespread distributed generation and storage devices. We have assumed that agents are free to trade electric energy over the network  and we have focused on the emerging market dynamics. We have considered three scenarios for the market dynamics for which we have presented a stability analysis. We have seen that stability depends on the specific form of the market dynamics and it may depend on the structure of the underlying network topology.  We have run numerical simulations that confirm our theoretical predictions. %A thorough study of Scenarios II and III is left for future investigations.

%We can extend the analysis in the previous section to consider that a situation in which energy is perceived as a rare commodity, whose price can quickly fluctuate in time. In this case, agents will have to adopt different strategies to cope with the fluctuations of the price of energy.
%We still consider Eq. (3), describing how the storage at each node $i$ evolves based on the prices set by the agents.
\bibliographystyle{IEEEtran}

\end{document}